\documentclass[twocolumn,tightenlines,showpacs,aps]{revtex4}
\usepackage{graphicx}

\begin{document}

\title{Temperature dependent anti-Stokes/Stokes ratios under Surface Enhanced Raman Scattering (SERS) conditions}

\author{R. C. Maher} \email{Robert.Maher@imperial.ac.uk}

\author{L. F. Cohen}

\author{J. C. Gallop}

\affiliation{The Blackett Laboratory, Imperial College London\\
Prince Consort Road, London SW7 2BW, United Kingdom}

\author{E. C. Le Ru}
\author{P. G. Etchegoin} \email{Pablo.Etchegoin@vuw.ac.nz}

\affiliation{The MacDiarmid Institute for Advanced Materials and Nanotechnology\\
School of Chemical and Physical Sciences, Victoria University of
Wellington,  PO Box 600 Wellington, New Zealand}

\date{\today}

\begin{abstract}
We make systematic measurements of Raman anti-Stokes/Stokes (aS/S)
ratios using two different laser excitations (514 and 633 nm) of
rhodamine 6G (RH6G) on dried Ag colloids over a wide range of
temperatures (100 to 350 K). We show that a temperature scan
allows the separation of the contributions to the aS/S ratios from
{\it resonance effects} and {\it heating/pumping}, thus decoupling
the two main aspects of the problem. The temperature rise is found
to be larger when employing the 633 nm laser. In addition, we find
evidence for mode specific vibrational pumping at higher laser
power densities. We analyze our results in the framework of
ongoing discussion on laser heating/pumping under SERS conditions.
\end{abstract}


\maketitle

\section{Introduction}

Despite rapid progress on Surface Enhanced Raman Scattering
(SERS)\cite{Moskovits,Otto} towards applications
\cite{SmithScreen,MinoScreen,MirkinDiog, KneippIntra}, many
fundamental questions remain open. An important issue --that has
been the subject of controversy-- relates to whether the local
enhanced electromagnetic (EM) fields cause local heating and/or
vibrational pumping. This is important for reasons that range from
the purely practical (damage to analytes under study) to the more
fundamental (vibrational pumping). In certain cases,
plasmon-related heating effects may even be
useful\cite{HalasTreat} for biomedical applications, but both the
magnitude of the effect and how to control it need proper
characterization. It is possible to use the aS/S ratio under SERS
conditions to study heating/pumping effects with certain
limitations, as we discuss in detail below.

The laser power dependence of the aS/S- ratio was thought to
provide strong evidence for SERS pumping \cite{Kneipp1}, but this
has been the source of controversy and alternative explanations
have been offered\cite{MCE1, Brolo, Haslett, HPdef}. Contributions
to the aS/S-ratio can be complex and include: heating, pumping,
and resonance effects. There are various hints for heating in SERS
and the anomalous ratio (commonly measured even at low powers) was
predicted to relate to hidden resonances in the
system\cite{Brolo,Haslett}. We have recently demonstrated that it
is indeed possible to use the aS/S ratio to map out roughly the
underlying resonance in certain cases\cite{MaherResonance}.

Here we develop the use of aS/S-ratios further; we expand and
provide additional evidence to recent discussions in Refs.
\cite{HPdef,RobFaraday}. We demonstrate how to separate heating
from resonance effects through temperature $(T)$ dependent data.
We find a temperature rise at low laser powers ($\approx$ 0.5 mW
on a $\sim$ 1.5 $\mu$m (diameter) spot) that depends on laser
energy. We find evidence also suggesting the existence of mode
specific vibrational pumping at higher laser power densities and
low temperature.

\section{General considerations}

For the sake of clarity we shall repeat the essential details of
the discussion presented in Refs.\cite{HPdef,RobFaraday} keeping
details to a minimum. We call $T_{env}$ the temperature of the
local environment (substrate + ambient surrounding the sample),
while $T_{room}$ is room temperature, $T_{met}$ the temperature of
the metal colloids, and $T_{mol}$ the temperature of the molecule.
The main possible scenarios are:

\begin{enumerate}
    \item No heating. The whole system is in thermal equilibrium; i.e. $T_{room}=T_{env}=T_{met}=T_{mol}$.
    \item Global heating: system in thermal equilibrium at a $T$ higher than $T_{room}$; i.e. $T_{env}=T_{met}=T_{mol}>T_{room}$.
    \item Local heating of substrate and analyte: when $T_{mol}=T_{met}\geq T_{env}\geq
    T_{room}$. The analyte is in thermal equilibrium with the
    metal, at a higher temperature than their environment. The
    source of heating here is mainly laser absorption in the metal
    substrate.
    \item Local heating of analyte $T_{mol}\geq T_{met}\geq T_{env}\geq T_{room}$.
    Here the molecule remains in internal thermal equilibrium at a
    temperature $T_{mol}$, higher than the rest of the system.
    Strong interactions among vibrational
    modes within the molecule allows vibrational energy to thermalize
    and reach an equilibrium through intramolecular vibrational
    relaxation or redistribution (IVR). The source of heating here
    can be direct laser absorption in the molecule (via electronic state
    transitions), or vibrational pumping (via SERS processes)
    followed by fast IVR. However, it is difficult to distinguish
    between these two possible causes.
    \item Mode specific vibrational pumping, when the analyte is not in internal thermal equilibrium.
    Each vibrational mode can be described by an effective mode
    temperature, $T^{eff}_i$, which reflects the population of its
    first vibrational excited state \cite{HPdef}. In general, IVR
    ensures that $T^{eff}_i$ is the same for all modes (internal thermal
    equilibrium at a temperature $T_{mol}$).
    However, SERS processes affect the vibrational population and
    can pump vibrations from the ground to the first excited
    vibrational state (for Stokes processes). If for a given vibrational mode, the pumping
    rate becomes comparable or higher than the IVR rate, this mode
    can then have a higher effective temperature than the other
    modes: $T^{eff}_i\geq T_{mol}\geq T_{met}\geq T_{env}\geq T_{room}$;
    this is referred to as {\it mode specific pumping}.
\end{enumerate}

The different scenarios deserve a brief comment. If vibrational
pumping occurs in the presence of fast IVR, pumping will only be
observed as an effective heating of the molecule (4$^{th}$
scenario). From the experimental point of view, under these
circumstances, it would be impossible to distinguish this
situation from the 3$^{rd}$ scenario. It is indeed very difficult
to determine independently the temperature of the metal and the
molecule, due to the lack of a spectroscopic feature coming
explicitly from the metal. Even if one could prove that the 4$^{th}$
scenario was the correct one, it would then be extremely difficult
to determine the real cause of heating: direct laser absorption or
vibrational pumping, or a combination of both effects.
Consequently, the only clear-cut demonstration of vibrational
pumping is, in our view, the last scenario; i.e. mode specific
pumping. Heating and pumping under SERS conditions have previously
been investigated using peak parameter correlations, such as width
or frequency vs. intensity\cite{HPdef,MCE2}. Evidence was found
for local heating (3$^{rd}$ or 4$^{th}$ scenario) but no evidence of
mode specific pumping was uncovered \cite{HPdef}. Although
yielding evidence for the existence of local heating, it was
impossible to distinguish its origin; i.e. either within the
molecule itself or the colloid (and consequently the molecule
through its close contact with the metal). These previous results
do not exclude the presence of pumping ({\it vide supra}) but they
do not demonstrate it either.

Most other studies of heating/pumping under SERS conditions
\cite{Kneipp1,MCE1, Brolo, Haslett} have focused on the
measurements of the aS/S ratio, $\rho$. This ratio is in principle
directly related to the vibrational populations, and therefore to
the effective mode temperature. However, in the case of SERS, it
can be modified by underlying resonances associated with the
wavelength dependence of the SERS enhancement factor, making the
interpretations more difficult. The most general expression for
the ratio under SERS conditions at low powers (in the absence of
pumping) and for molecules in thermal equilibrium at a temperature
$T_{mol}$) is\cite{Kneipp_rev,Xu2}:


\begin{equation}
\rho = \frac{I_{AS}}{I_S} = A_\omega
\frac{\sigma_{AS}}{\sigma_S}\sum_{i=1}^N\frac{A_{AS}}{A_S}\exp(-\hbar\omega_v/k_B
T_{mol}) \label{ratio-corrected}
\end{equation}
with
$A_\omega=(\omega_{L}+\omega_{v})^4/(\omega_{L}-\omega_{v})^4$
(accounting for the standard wavelength dependence of Raman
processes). $I_{AS(S)}$ is the intensity of the anti-Stokes
(Stokes) Raman mode, $\omega_{L(v)}$ is the frequency of the laser
(Raman mode), $\sigma_{AS(S)}$ is the anti-Stokes (Stokes)
scattering cross section of the molecule, $N$ is the number of
active molecules, and $A_{AS(S)}$ is the SERS enhancement factor
at the molecule position at the anti-Stokes(Stokes) frequency. The
latter term includes both electromagnetic and chemical enhancement
effects. In general the enhancement can be site dependent and
therefore should be summed over all active sites. Because
anti-Stokes signals are typically small, their measurement
requires relatively long integration time and/or large analyte
concentrations. The results are therefore an average over several
active sites/molecules, each with different enhancement factors
and resonances. To account for this averaging, $\rho$ for a given
mode $i$ can be modelled in a first approximation as:
\begin{equation}
\rho_i=A_\omega A_i \exp(-\hbar\omega_i/k_B T_i^{eff})
 \label{ratioAT}
\end{equation}
where $A_i$ is the average {\it asymmetry factor} from resonance
effects (asymmetries in cross sections), which can change from
mode to mode. $T_i^{eff}$ was used in this expression to account
for the possibility of different effective mode temperatures. When
the vibrational pumping rate is small compared to IVR rate, we have
$T_i^{eff}=T_{mol}$ for all modes and the molecule is in thermal
equilibrium at $T_{mol}$. However, if vibrational pumping is not
negligible, this expression needs to be modified to give (for weak
pumping) \cite{Kneipp1, Haslett,HPdef}:
\begin{equation}
\rho_i=A_\omega A_i \left[\exp(-\hbar\omega_i/k_B
T_{mol})+\sigma_S \tau_r I_L \right]
 \label{ratioATI}
\end{equation}
where $\tau_r$ is the vibrational relaxation time (including IVR
and other relaxation processes), and $I_L$ is the laser intensity
in photons per unit time and surface area. We see in this latest
expression that vibrational pumping would tend to increase the
value of $\rho_i$. This could alternatively be described by Eq.
(\ref{ratioAT}), using an effective mode temperature
$T_i^{eff}>T_{mol}$, but determining $T_i^{eff}$ would require the use of Eq. (\ref{ratioATI}).

Eqs. (\ref{ratioAT}) or (\ref{ratioATI}) form the basis for the
analysis of most SERS experiments about heating/pumping that use
aS/S ratio measurements. From them, it is obvious that the
measurement of $\rho_i$ only is not sufficient to extract
$T_{mol}$ or $T_i^{eff}$, as the $A_i$'s are unknown. One
therefore needs an extra free parameter. In previous works, the
power dependence of $\rho_i$ was for example studied
\cite{Kneipp1,Haslett,MCE2}. A linear dependence of $\rho_i$ with
$I_L$ has been invoked as evidence of vibrational pumping which
would appear to follow from Eq.\ (\ref{ratioATI}). This is not true
however and direct heating of the molecule has been proposed as an
alternative explanation \cite{Brolo}. To understand the effect of
heating on $\rho$, one can write $T_{mol}=T_0+\Delta T$, and
expand Eq. (\ref{ratioAT}) considering a small temperature
increase, $\Delta T \ll T_0$:
\begin{equation}
\rho_i=A_\omega A_i \exp \left( -\frac{\hbar\omega_i}{k_B T_0}
\right) \exp \left( \frac{\hbar\omega_i}{k_B T_0} \frac{\Delta
T}{T_0} \right).
 \label{ratioheating}
\end{equation}
If heating effects are present, $\Delta T$ changes with power.
Because heat diffusion and transfer models are linear, it is
reasonable to assume that the temperature increase $\Delta T$ is
proportional to laser power: $\Delta T = a I_L$. For a small
temperature increase, the argument in the second exponential of
Eq. (\ref{ratioheating}) is small and expanding it leads to
\begin{equation}
\rho_i=A_\omega A_i \exp \left( - \frac{\hbar\omega_i}{k_B T_0}
\right) \left( 1 + \frac{\hbar\omega_i}{k_B T_0} \frac{a}{T_0} I_L
\right).
 \label{ratioheating2}
\end{equation}
The linear dependence of $\rho_i$ (or quadratic dependence of
$I_{AS}$) with incident power can therefore equally be the result of conventional
heating effects or of pumping. It is also worth noting that the
argument in the second exponential of Eq. (\ref{ratioheating}) can
be comparable to 1, even for a small $\Delta T$. This can be the
case for example for high energy modes, where $\hbar\omega_i / k_B
T_0$ can be $\approx$ 10. The power dependence of $\rho_i$ as a
result of heating effects would then be exponential. Such power
dependence, linear for low energy modes, and exponential for high
energy modes, has already been observed \cite{MCE2}. This shows
that heating effects can be important and that a linear power
dependence of $\rho_i$ is not in itself a proof of vibrational
pumping.

Instead of using the laser power as an additional parameter, we propose to study the aS/S ratios of each mode as a function
of substrate temperature $T_0$. As we shall show, by investigating
how these ratios change as a function of $T$ over a wide range, we
can separate the contributions due to heating and resonance
effects. In addition, this approach enables us to detect a departure
from internal thermal equilibrium of the molecule, i.e.
vibrational pumping.

\section{Experimental details}

Silver colloids were produced as described in Ref.\cite{Miesel}.
SERS samples were prepared by mixing equal amounts of colloidal and 25 mM KCl solutions
together, to which 1 $\mu$L of 10$^{-6}$ M
Rhodamine 6G (RH6G) solution was added, resulting in a 10$^{-9}$ M
RH6G concentration. The sample was then dried onto a Si substrate.
The temperature calibration in terms of aS/S-data of non-SERS
active signals (paracetamol) is described
elsewhere\cite{RobFaraday}. Measurements were carried out with a
Renishaw 2000 CCD spectrometer and an Olympus BH-2 microscope with
a Linkham temperature stage, using both 514 nm Ar$^{+}$ ion and
633 nm HeNe lasers with a $\times$50 long working distance
objective ($\sim$1.5 $\mu$m beam diameter at the focal point).
Measurements with the 514 nm laser were made at 0.5 mW whereas for
the 633nm laser were performed at 0.5 and 5 mW, respectively, in
the range between 100 and 350 K in steps of 10 K and integration
times varying from 1 to 30 sec. Five or more measurements were
made to gain an average over the sample. Peaks were then analyzed
using standard Voigt functions with subtracted backgrounds.

\begin{table*}
    \centering
        \begin{tabular}{|c|c|c|c|c|c|c|}
          \hline
       & A + $\Delta T$ shared & \multicolumn{2}{|c|}{(A + $\Delta T$)} & A only & $\Delta T$ only \\
          \hline
            Mode (cm$^{-1}$) & A & A & $\Delta T$ & A & $\Delta T$ \\
            \hline
            \multicolumn{6}{|c|}{Low power 633 nm laser} \\
            \hline
            610 & 1.30 $\pm$ 0.05 ($\Delta T$ = 25K) & 1.50 $\pm$ 0.05 & 17 $\pm$ 5 & 2.00 $\pm$ 0.20 & 40 $\pm$ 5 \\
            \hline
            780 & 1.35 $\pm$ 0.05 ($\Delta T$ = 25K) & 1.65 $\pm$ 0.10 & 16 $\pm$ 5 & 2.35 $\pm$ 0.20 & 35 $\pm$ 5 \\
            \hline
            1310 & 0.85 $\pm$ 0.10 ($\Delta T$ = 25K) & 0.50 $\pm$ 0.10 & 55 $\pm$ 10 & 1.60 $\pm$ 0.30 & 20 $\pm$ 5 \\
            \hline
            1360 & 0.95 $\pm$ 0.10 ($\Delta T$ = 25K) & 0.80 $\pm$ 0.10 & 35 $\pm$ 5 & 1.90 $\pm$ 0.30 & 25 $\pm$ 5 \\
            \hline
            1510 & 1.05 $\pm$ 0.10 ($\Delta T$ = 25K) & 0.70 $\pm$ 0.10 & 40 $\pm$ 6 & 2.20 $\pm$ 0.40 & 30 $\pm$ 10 \\
            \hline
            \multicolumn{6}{|c|}{High power 633 nm laser} \\
            \hline
            610 & 0.45 $\pm$ 0.05 ($\Delta T$ = 96K) & 1.05 $\pm$ 0.10 & 40 $\pm$ 5 & 3.1 $\pm$ 0.5 & 40 $\pm$ 5 \\
            \hline
            780 & 0.40 $\pm$ 0.04 ($\Delta T$ = 96K) & 0.70 $\pm$ 0.10 & 60 $\pm$ 15 & 4.4 $\pm$ 1 & 50 $\pm$ 5 \\
            \hline
            1310 & 0.50 $\pm$ 0.10 ($\Delta T$ = 96K) & 0.15 $\pm$ 0.05 & 160 $\pm$ 35 & 12 $\pm$ 3 & 75 $\pm$ 10 \\
            \hline
            1360 & 0.40 $\pm$ 0.10 ($\Delta T$ = 96K) & 0.25 $\pm$ 0.10 & 120 $\pm$ 35 & 14 $\pm$ 3 & 70 $\pm$ 10 \\
            \hline
            1510 & 0.50 $\pm$ 0.10 ($\Delta T$ = 96K) & 0.08 $\pm$ 0.03 & 200 $\pm$ 40 & 20 $\pm$ 6 & 80 $\pm$ 10 \\
            \hline
            \multicolumn{6}{|c|}{Low power 514 nm laser} \\
            \hline
            610 & & 0.45 $\pm$ 0.05 & 3 $\pm$ 2 & 0.50 $\pm$ 0.03 & -20 $\pm$ 4 \\
            \hline
        \end{tabular}
    \caption{Fitting results for the 514 and 633 nm lasers using the models described in the text.
    Results are shown for both high and low power measurements for the 633 nm laser whilst only low power measurements were
    possible with the 514 nm laser.  Low power means  $\sim$0.5 mW at the focal point whilst the high power measurement
    was an order of magnitude larger.  All measurements were made on large colloidal
    clusters.The $A$+$\Delta T$-shared fitting is not possible for the 514 nm laser as only one mode was
    measurable. The ($A$+$\Delta T$-shared) and ($A$+$\Delta T$) models provide the best representation of the data
    at low and high power respectively.}
    \label{tab:FitRes633a}
\end{table*}

The variation in the aS/S ratio for each mode was fitted to Eq.
(\ref{ratioAT}), with $T_i^{eff}=T_{Si}+\Delta T_i$., where
$\Delta T_i$ is the increase in effective mode temperature with
respect to the $T$ of the substrate which is measured by the
aS/S-ratio of the Si substrate\cite{RobFaraday}. The parameters
$A_i$ in Eq. (\ref{ratioAT}) will depend on the particular mode
under consideration, accounting for the asymmetry in $\sigma_S$
and $\sigma_{aS}$ produced by resonances. Four types of fits were
performed:
\begin{itemize}
 \item $\Delta T$-only,
  where $A_i$ is set to 1 (no resonance
effect) and $\Delta T_i$ are fitting parameters;
 \item $A$-only,
  where $\Delta T_i=0$ (no local heating, no pumping) and $A_i$
  are fitting parameters;
  \item $A$+$\Delta T$-shared,
  where $\Delta T_i= \Delta T$ constant for all modes and $A_i$
  are fitting parameters. This corresponds to the case of internal
   thermal equilibrium for the molecule;
   \item $A$+$\Delta T$,
   where $A_i$ and $\Delta T_i$ are independent fitting
   parameters. This allows the modes to have different effective
   mode temperatures.
 \end{itemize}

\begin{figure}
 \centering{
  \includegraphics[width=9cm, height = 12cm]{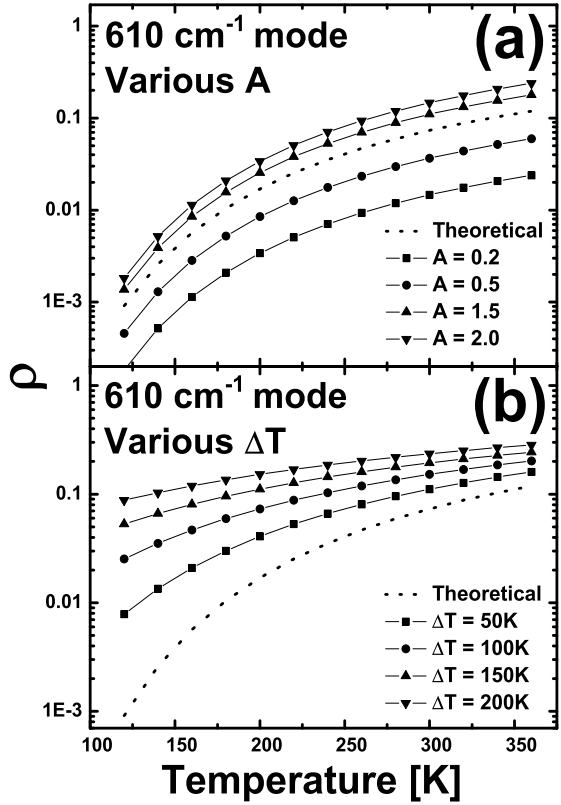}
 }
\caption{Theoretical plots showing the effect on the
anti-Stokes/Stokes ratio of various values of $A$ (left) and
$\Delta T$ (right).  Differing values of $A$ result in nested
anti-Stokes/Stokes ratios over the full temperature range.  In the
case of $\Delta T$, the effect on the anti-Stokes/Stokes ratio
varies across the full temperature range with the largest effect
being made at lower temperatures where $\Delta T$ represents a
greater percentage of the initial temperature.  This results in a
`flaring out' of the ratio at lower temperatures which is
noticeably larger for greater values of $\Delta T$.}
 \label{TheoADT}
\end{figure}

Figure \ref{TheoADT} illustrates the predictions of the model and
the influence of $A$ and $\Delta T$ on the value of the aS/S-ratio
as a function of $T$. Figure \ref{TheoADT}(a) shows the expected
value of the ratio for different $A$'s, whilst keeping $\Delta
T=0$. When $A<1$ ($A>1$) the ratio is suppressed (increased). In
Fig.\ref{TheoADT}(b) $A$ is held constant whilst $\Delta T$ is
varied: the influence of $\Delta T$ is strongest at low $T$'s
where it is a greater fraction of the initial $T$, resulting in a
distinctive `flaring' of the ratio at low $T$'s and a flattening
of the overall curve shape.

The first two types of fits are only used to test whether a simple
model with heating only and no resonances, or resonances only and
no heating, is sufficient to explain the data. For all fits, we
assume that $A_i$ and $\Delta T_i$ are constant with temperature.
$A_i$, which is the result of predominantly electromagnetic
resonances, is not believed to vary with temperature. For
molecules in thermal equilibrium $\Delta T_i= \Delta T$ is the
same for all modes and represent the temperature increase
(heating). Many factors contribute to this increase, for example
the thermal properties of the SERS substrate and of the various
interfaces. These factors could possibly vary with temperature,
although large variations typically occur mainly at even lower
temperatures ($T<100$\,K). The assumption $\Delta T=$constant is
therefore reasonable in a first approximation. The ($A$+$\Delta
T$-shared) fit then enables us to separate the resonance effects
($A_i$) from the heating effect ($\Delta T$). If this fit fails,
then either $\Delta T=$constant is wrong, or the molecule is not in
internal thermal equilibrium (i.e. pumping occurs). In the latter
case, the effective mode temperature $T_0+\Delta T_i$ is
determined by the pumping rate, via Eq. (\ref{ratioATI}). The fits
with this model cannot therefore be valid, but the ($A$+$\Delta
T$) fit can still be useful in identifying which modes present an
anomalous effective mode temperature and are therefore most likely
to be pumped.

If the ($A$+$\Delta T$-shared) fit fails, the model above cannot
distinguish directly between pumping ($\Delta T_i$ different for
different modes) and temperature dependent heating ($\Delta
T_i=\Delta T$, but $\Delta T$ varies with $T$). In order to prove
that different modes cannot be accounted for by a single $T$, we
propose to define the following function (for two modes $a$ and
$b$):
\begin{equation}
F\equiv \frac{k_B{\rm ln}(\rho_a)}{\hbar\omega_a}- \frac{k_B{\rm
ln}(\rho_b)}{\hbar\omega_b}.
 \label{F1}
\end{equation}
Then, describing each mode by an effective mode temperature, we
have from Eq. (\ref{ratioAT}).
\begin{equation}
F\equiv C+\left[\frac{1}{T_a^{eff}}-\frac{1}{T_b^{eff}}\right];
 \label{F}
\end{equation}
where $C$ is a constant accounting for the different A-factors.
The advantage of this function is as follows: if both modes share
the same temperature, this function {\it is constant as a function
of $T$}. The function $F$ can be used to monitor relative changes
between the effective $T$'s of two modes as a function of external
varying conditions (here $T$, but it could also be used to prove
non-equilibrium pumping in a power dependence experiment).

\section{Results and discussion}

\subsection{Low power}

\begin{figure}
 \centering{
  \includegraphics[width=9cm, height = 12cm]{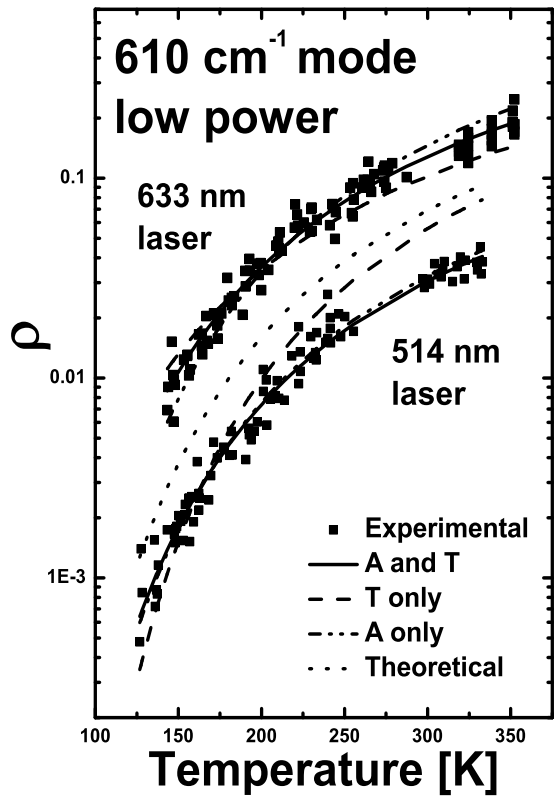}
 }
\caption{Anti-Stokes/Stokes ratios as a function of temperature
for the 610 cm$^{-1}$ SERS active modes of RH6G for the 633 nm and
514 nm lasers at low power (0.5 mW). There is a clear different in
the resonance contribution to the anomalous ratios between these
lasers as the 633 nm ratio is increased above the theoretical
value whilst the 514 nm ratio is suppressed below the theoretical.
The solid lines represent the fit to the experimental data using
the ($A$+$\Delta T$) model whilst the dashed and dash dot
represent the fits obtained from using only the $\Delta T$ or $A$
parameter respectively.  The dotted line is the theoretical ratio
values. See text for details.}
 \label{633514nm610lo}
\end{figure}

Figure \ref{633514nm610lo} shows examples aS/S ratios as a
function of $T$ for the 610 cm$^{-1}$ mode of RH6G and 633 nm
excitation (top curve). The theoretically predicted behavior
($A$=1, $\Delta T=0$) and the fits to the experimental data using
the ($A$+$\Delta T$), $A$-only, and $\Delta T$-only models, are
shown. The experimental data are well approximated by the
($A$+$\Delta T$-shared) fit for all modes investigated. The
fitting to all the modes, using the four fitting procedures
presented above, is summarized in table \ref{tab:FitRes633a}.
Figure \ref{633514nm610lo} also shows the equivalent data for the
610 cm$^{-1}$ mode under the 514 nm laser excitation (bottom
curve). The aS-peaks associated with the higher modes were too
small to analyze at these laser powers.  The (A + $\Delta T$)
fitting provides the best fit to the experimental data although
$\Delta T$ is small and consequently the A only model also fits
the data reasonably well. The fitting results for the 610
cm$^{-1}$ mode taken using the 514 nm laser are also summarized in
table \ref{tab:FitRes633a}.

The low power measurements tell us that  the molecule is in thermal equilibrium with a heating of ~25K when using 633nm laser and 3K when using the same low power 514nm laser. Useful insight into the colloidal heating problem can be gained from considering a simplified 1-D model. Consider a single colloid where the laser power is assumed to dissipate through the top surface with no lateral heat flow.  The $T$ dependence of the thermal conductivity of Ag is small down to 40 K and we can use the room temperature value. The internal thermal time constant of the colloid (the time for the colloid to come to a uniform temperature), $\tau_c$ is given by $\tau_c = \frac{C_{Ag}}{K_c}\approx 1\times10^{-11}(s)$, where $K_c$ is the thermal conductance of a single Ag particle (of size $\sim (20 \times 10^{-9}$ m)$^3$) and $C_{Ag}$ is the phononic contribution to the specific heat with Debye temperature, $\Theta_{Ag}$ given by: $C_{Ag}=c_{Ag}\left((D_{Ag}/MW_{Ag}\times10^{-3})\times d^3 \right)\approx 6.1\times10^{-17}(J/K)$. In this latter expression $W_{Ag}$ is the atomic weight of Ag, $D_{Ag}$ its density, $d$ the length of the colloid and $c_{Ag}\approx 7.8\times10^1(J/Mol.K)$. We ignore the electronic contribution to $c_{Ag}$ as it is much smaller than the phonon contribution, except at liquid He temperatures.

Assuming an incident power 1 mW is focussed into a 1 $\mu$m$^2$ area with a reflectivity of 80\%, the number of photons absorbed per second by the colloid is $\approx 2.6 \times 10^{11}$s$^{-1}$. Note that when dealing with small objects, $\tau_c$ is comparable to the arrival rate of the photons. The dynamics of the system is therefore rather interesting. The absorption of a single photon increases the colloid's upper surface temperature by $\approx 5.2 \times 10^{-3}$ K.  By modelling the acoustic mismatch between Ag and Si\cite{modelone}, we can estimate the expected increase in $T$.  Estimating values from experimental data for interfaces between Al-Al$_2$O$_3$ and Pb-diamond, we find temperature increases of 2 and 10 K, respectively, which are the of the right order of magnitude for our low power observations using the 633nm laser. This is remarkable when we consider that we have ignored any potential contribution originating from SERS hot-spots and that the thermal model is extremely simplified. This model supports the scenario that  at low powers the molecules could be heating through direct interaction with the colloids.

\subsection{High power}

\begin{figure}
 \centering{
  \includegraphics[width=9cm, height = 12cm]{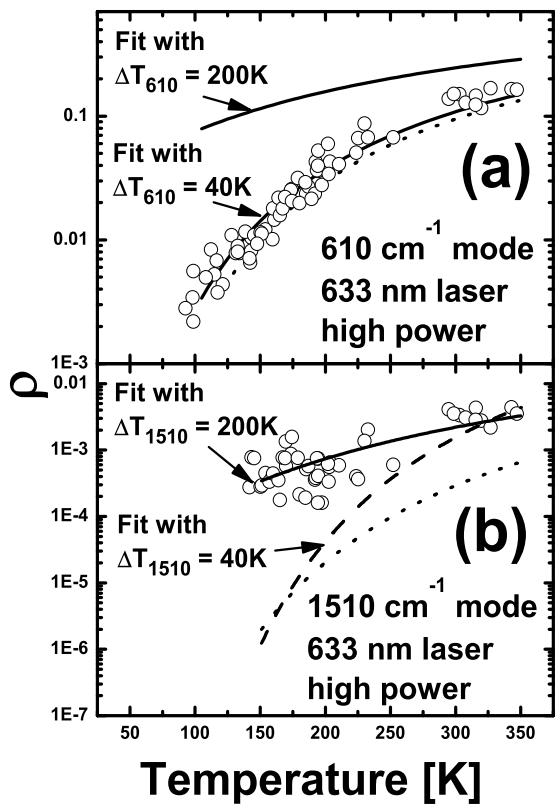}
 }
\caption{Anti-Stokes/Stokes ratio as a function of temperature for
the 610 and 1510 cm$^{-1}$ SERS active modes of RH6G taken with
the 633 nm laser at high power (5 mW).  The theoretically expected
ratio is shown as the dotted line.  Two fits with the same value
of $A$ are shown for each mode.  $\Delta T_{610}$ = 40K and
$\Delta T_{1510}$ = 200K represents the best fit values for these
modes. The secondary fits set $\Delta T_{610}$ = 200K and $\Delta
T_{1510}$ = 40K.}
 \label{633nmhi}
\end{figure}

Figure \ref{633nmhi} shows $\rho$ as a function of $T$ for the 610
and 1510 cm$^{-1}$ modes for high power (5 mW) 633 nm excitation,
while all other parameters were identical to the measurements at
low power. We show the resulting fits of ($A$+$\Delta T$) for both
modes, as well as secondary fits swapping the temperatures and the theoretically expected $\rho$'s. The high power ratios for
all the modes were also fitted using the ($A$+$\Delta T$-shared)
model as in the low power data.  This resulted in a $\Delta
T=96$\,K whereas the low power results gave 25\,K. Although this
is clearly larger, as would be expected, the quality of the fits
were badly compromised for the high power data, as measured by a
$\chi^2$-test (not shown in the table). This indicates that this
model is no longer valid in this regime. The results of the
fitting to the high power, 633 nm data for all the modes is also
summarized in table \ref{tab:FitRes633a}.
 The 610 cm$^{-1}$ ratio
as a function of $T$ at high power is similar to that obtained at
low powers; when plotted on the same graph it is almost impossible
to separate them by eye suggesting that the mode temperature is
not too dissimilar, despite an order of magnitude difference in
the incident power. Indeed, the result of the ($A$+$\Delta T$)
fitting of the 610 cm$^{-1}$ mode in both cases gives $\Delta T$ =
$17 \pm 5$ K for low power and $\Delta T$ = $40 \pm 5$ K for high
power, which is difficult to visualize on a log-scale. For the
1510 cm$^{-1}$ mode, although there is a great deal of scatter,
there is a noticeable difference in the trend of the data when
compared with the low power measurement. This behavior is very
similar to the `flaring' of the theoretically determined $\rho$'s
for high $\Delta T$ as shown in figure \ref{TheoADT}. This is
confirmed with the fitting of the data to the (A+$\Delta T$) model
which results in $\Delta T_{1510}$ = $200 \pm 40$ K; i.e. much
larger than the $T_{610}$ = $40 \pm 5$ K from the best fit to the
610 cm$^{-1}$ mode at high power. Figure \ref{633nmhi} also  shows fits
for the 610 and 1510 cm$^{-1}$ modes with $\Delta T_{610} = $200K
and $\Delta T_{1510} = 40$, both with the best fit value of A for
that mode.  It is obvious from these fits that the 610 cm$^{-1}$
mode is much cooler than the 1510 cm$^{-1}$. All these results
suggest that vibrational pumping in the 1510 cm$^{-1}$ mode could
be the explanation for the failure of the ($A$+$\Delta T$-shared)
fits.

\begin{figure}
 \centering{
  \includegraphics[width=10cm, height = 10cm]{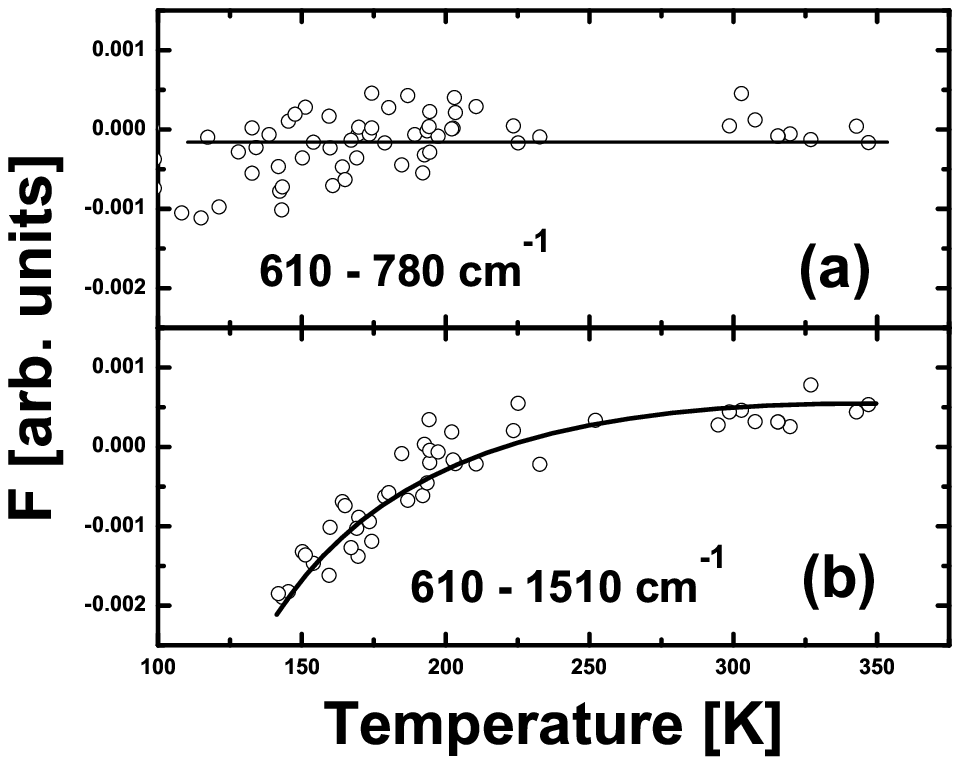}
 }
\caption{$F$ function (Eq. \ref{F}) vs. $T$ comparing the
effective temperatures of two different pairs of modes: (a) 780
and 610 cm$^{-1}$ and (b) 1510 and 610 cm$^{-1}$. A constant value
of $F$ vs. $T$ in (a) means that these two modes share a common
temperature. The comparison in (b) however points towards mode
selective pumping of the 1510 cm$^{-1}$ mode with respect to the
610 cm$^{-1}$. Solid lines are fits to the data. See the text for
further details.}
 \label{fig4}
\end{figure}

In order to confirm this hypothesis, we plot in Fig. \ref{fig4},
the function $F$, discussed in the previous section, which
compares the effective $T$'s of (a) the 780 and 610 cm$^{-1}$
modes and (b) the 1510 and 610 cm$^{-1}$ modes in (b). The
monotonic departure from a constant observed in (b) is an
additional demonstration of non thermal equilibrium between these
two modes and therefore in the molecule. This confirms that the
failure of the ($A$+$\Delta T$-shared) fits is a consequence of
vibrational pumping, and not of a temperature dependent heating
(same $\Delta T$ for all modes, but varying with temperature). In
fact, the data in Fig. \ref{fig4} show not only evidence for mode
specific pumping but also highlights the limitations of the
$A$+$\Delta T$ model. The kink in $F$ as a function of $T$ means
that the 1510, 780, and 610 cm$^{-1}$ modes share a single $T$
above $\sim$ 200 K, whilst below that the 1510 cm$^{-1}$ shows
some mode specific pumping. This is only taken into account
approximately in the $A$+$\Delta T$ model, and Eq.
(\ref{ratioATI}) needs to be used to account for pumping properly.
Fits using Eqs. (\ref{ratioATI}) and (\ref{F1}) are also shown in
Fig. \ref{fig4} and are in good agreement with the data. Moreover,
we extract from the fit for the 1510 cm$^{-1}$ mode a value of
$\sigma_S \tau_r I_L = 5\times 10^{-5}$. We estimate $I_L \approx
10^{24}$\,photons/cm$^2$ and typical values for $\tau_r$ are in
the range 1--10\,ps. This leads to $\sigma_S$ in the range
$5\times 10^{-16} - 5\times 10^{-15}$, compatible with previous
estimate of the SERS cross section. The study of the
$T$-dependence of the function $F$ therefore provide additional
evidence that the behavior of the 1510 cm$^{-1}$ below 200\,K is a
manifestation of vibrational pumping in this mode. This effect is
more prominent at low $T$ and for higher energy modes because the
equilibrium population of the excited vibrational state is then
very small and the relative change in intensity is larger.
Following this argument, further evidence for pumping could be
obtained from measurements at even lower temperatures ($T < 100$K)
than the ones accessed in here. We note that the original paper on
SERS pumping also presented some evidence for mode dependent
effects\cite{Kneipp1}. Our results certainly contribute to this
body of work and suggest that several mechanisms (vibrational
pumping, underlying resonances, and colloid or SERS substrate
heating) operate simultaneously and contribute together to the
aS/S ratio anomalies.

In summary, we detail the key results presented in
table \ref{tab:FitRes633a} as follows: $(i)$ The ($A$+$\Delta
T$-shared) model provides a satisfying agreement at low powers.
The molecule is then in thermal equilibrium with a heating of
$\sim 25$ K with respect to the substrate for the 633 nm laser,
but only a 3 K rise for 514 nm excitation. $(ii)$ Models with only
A or $\Delta T$ alone as fitting parameters systematically produce
worse overall agreement with the data or even unphysical values,
like negative heating (see for example the 610 cm$^{-1}$ mode at
514 nm excitation in table \ref{tab:FitRes633a}).
 $(iii)$ For high powers
using the 633 nm laser, an overall shared $\Delta T$ among the
modes is not possible and one is forced to a better quality fit by
allowing different effective temperatures for different modes.
Using the function $F$, one can then demonstrate that this is the
result of vibrational pumping in the 1510 cm$^{-1}$ mode.

\section{Conclusion}

In closing, we have observed a laser frequency dependent heating
of the order of 25 K at low laser power under SERS conditions with
633 nm excitation. We have shown that it is possible to decouple
in dry samples the contributions of the thermal population and
resonances in the aS/S ratios by performing a $T$-scan, thus
overcoming the basic uncertainty in the different relative
contributions which exist in studies at a single $T$. Most
importantly we have used this technique to provide evidence for at
least two possible scenarios: molecule in thermal equilibrium at
low powers and mode specific pumping at high powers. Experiments
at lower temperatures are in progress and will be reported
elsewhere.

PGE and LFC acknowledge support by EPSRC (UK) under grant
GR/T06124. RCM acknowledges partial support from the National
Physical Laboratory (UK) and the hospitality of the MacDiarmid
Institute at Victoria University (NZ).

\newpage

\newpage

\newpage

\newpage


\begin{thebibliography}{}
\bibitem{Moskovits} M. Moskovits, Rev. Mod. Phys. {\bf 57}, 783 (1985).
\bibitem{Otto} A. Otto, in {\it Light Scattering in Solids}, edited by M. Cardona and G. G\"untherodt (Springer, Berlin, 1984), p. 289.
\bibitem{SmithScreen} K. Faulds, R. P. Barbagallo, J. T. Keer, W. E. Smith, D. Graham, Analyst {\bf 129}, 567 (2004).
\bibitem{MinoScreen} M. Green, F. M. Lui, L. F. Cohen, P. Kollensperger, and A. Cass, In
Press.
\bibitem{MirkinDiog} J. J. Storhofff, R. Elghanian, C. A. Mirkin, R. L. Letsinger, Langmuir {\bf 18}, 6666 (2002).
\bibitem{KneippIntra} J. Kneipp, H. Kneipp, W. L. Rice, and K. Kneipp, Anal. Chem. {\bf 77}, 2381 (2005).
\bibitem{HalasTreat} C. Loo, A. Lowery, N. Halas, J. West, R. Drezek, Nano Letters, {\bf 5}, 709 (2005).
\bibitem{Kneipp1} K. Kneipp, Y. Wang, H. Kneipp, et al., Phys. Rev. Lett. {\bf 76}, 2444 (1996).
\bibitem{MCE1} R. C. Maher, L. F. Cohen, P. Etchegoin, H. J. N. Hartigan, R. J. C. Brown, and M. J. T. Milton, J. Chem. Phys. {\bf 120}, 11746 (2004).
\bibitem{Haslett} T. L. Haslett, L. Tay, and M. Moskovits, J. Chem. Phys. {\bf 113}, 1641 (2000).
\bibitem{Brolo} A. G. Brolo, A. C. Sanderson, and  A. P. Smith, Phys. Rev. B {\bf 69}, 045424 (2004).
\bibitem{HPdef} E. C. Le Ru and P. G. Etchegoin, Faraday Discussions {\bf 132} (in press).
\bibitem{MaherResonance} R. C. Maher, J. Hou, L. F. Cohen, F. M. Liu, N. Green, R. J. C. Brown, M. J. T. Milton, E. C. Le Ru, J. M. Hadfield, J. E. Harvery, and P. G. Etchegoin, J. Chem. Phys. {\bf 123}, 084702 (2005).
\bibitem{RobFaraday} R. C. Maher, L. F. Cohen, E. C. Le Ru, and P. G. Etchegoin, Faraday Discussions {\bf 132} (in press).
\bibitem{MCE2} R. C. Maher, M. Dalley, E. C. Le Ru, L. F. Cohen, P. G. Etchegoin, H. Hartigan, R. J. C. Brown, and M. J. T. Milton, J. Chem. Phys. {\bf 121}, 8901 (2004).
\bibitem{Kneipp_rev} K. Kneipp, H. Kneipp, I. Itzkan, R. R. Dasari, and M. S. Feld, J. Phys: Condens. Matter {\bf 14}, R597 (2002).
\bibitem{Xu2} H. Xu, J. Aizpurua, M. K\"all, and P. Apell,  Phys. Rev. E {\bf 62}, 4318 (2000).
\bibitem{Miesel}P.C. Lee, D. Meisel, J. Phys. Chem. {\bf 86}, 3391 (1982).
\bibitem{modelone} D. G. Cahill, W. K. Ford, K. E. Goodson, G. D. Mahan, A. Majumdar, H. J. Maris, R. Merlin, S. R.
Phillpot, J. Appl. Phys. {\bf 93}, 793 (2003).
\end{thebibliography}
\end{document}